\documentclass[11pt,twoside,a4paper]{article}
\usepackage{fancyhdr}
\usepackage{url}
\newcommand{\cancel}[1]{}

\usepackage{cite}
\usepackage{amsmath,amssymb,amsfonts}
\usepackage{algorithmic}
\usepackage{graphicx}
\usepackage{textcomp}
\usepackage{makecell}
\usepackage{hyperref}
\usepackage{url}

\usepackage{multirow}
\usepackage{caption,setspace}

\def\BibTeX{{\rm B\kern-.05em{\sc i\kern-.025em b}\kern-.08em
    T\kern-.1667em\lower.7ex\hbox{E}\kern-.125emX}}

\newcommand{\shorten}[1]{}

\begin{document}

\title{VerifyMed - A blockchain platform for transparent trust in virtualized healthcare: Proof-of-concept}

\author{Jens-Andreas Hanssen Rensaa\thanks{Department of Information Security and Communication Technologies, Norwegian University of Science and Technology - NTNU} \and Danilo Gligoroski$^*$ \and Katina Kralevska$^*$ \and Anton Hasselgren\thanks{Department of Neuromedicine and Movement Science, Norwegian University of Science and Technology - NTNU} \and Arild Faxvaag$^\dagger$}

\maketitle

\begin{abstract}
Patients living in a digitized world can now interact with medical professionals through online services such as chat applications, video conferencing or indirectly through consulting services. These applications need to tackle several fundamental trust issues: 1. Checking and confirming that the person they are interacting with is a real person; 2. Validating that the healthcare professional has competence within the field in question; and 3. Confirming that the healthcare professional has a valid license to practice. In this paper, we present VerifyMed - the first proof-of-concept platform, built on Ethereum, for transparently validating the authorization and competence of medical professionals using blockchain technology. Our platform models trust relationships within the healthcare industry to validate professional clinical authorization. Furthermore, it enables a healthcare professional to build a portfolio of real-life work experience and further validates the competence by storing outcome metrics reported by the patients. The extensive realistic simulations show that with our platform, an average cost for creating a smart contract for a treatment and getting it approved is around 1 USD, and the cost for evaluating a treatment is around 50 cents.
\end{abstract}

\newpage
\tableofcontents

\newpage
\section{Introduction}
The healthcare industry is currently ongoing through a digital transformation, and innovations within information, and communication technologies have enabled the healthcare industry to improve the delivery of health services. Similar to Industry 4.0, Healthcare 4.0  \cite{Thuemmler2017} aims to use modern digital technologies to enable a virtualized healthcare environment by providing distributed and patient-centered care delivery. This virtualization is expected to accelerate with the emergence of next-generation mobile network strategies (5G) and artificial intelligence (AI), enabling virtualized care services to be executed in real-time and performed based on real-time data collection from anywhere at any time. 

The transition to virtualized health services poses some challenges; one of them is providing trust. In the current healthcare environment, most patients meet physically with healthcare workers in accredited healthcare institutions. However, when the meeting is moved to a virtualized environment, this inherited trust is decreased. Furthermore, building up such a trust relationship is even harder if the caregiver is an AI health worker. The health domain, therefore, needs new solutions to enable an establishment of trust between patients and healthcare workers in a virtualized environment.

Blockchain is a maturing technology with properties that can provide trust within a virtualized health domain as it allows mutually mistrusting entities to interact without the presence of a central trusted third party. While initially intended for the financial domain, the addition of smart contracts allow for general purpose applications to be made.
By creating and deploying smart contracts, we can build models that capture the authorization and the experience of a healthcare worker directly on the blockchain. These models can then be used to establish trust in a patient-caregiver encounter.

When it comes to building network applications that capture and nourish the complex relations of trust among different entities, a related area is the area of distributed database systems. As described in \cite{raikwar2020trends}, in the last decade, we witnessed a mutual influence and development between database technology and the blockchain technology. In particular, the blockchain technology has influenced the introduction of new functionalities in some modern databases such as immutability, privacy, and censorship resistance. Those blockchain functionalities are precisely the ones that we valued the most in this work.

\textbf{Our contribution:} We describe the design rationale, implementation and evaluation of VerifyMed - a proof-of-concept for transparently validating the authorization and competence of healthcare workers by using blockchain technology\footnote{As an online addition to this article, the source code and instructions for setting up the platform are available at \url{https://github.com/jarensaa/transparent-healthcare}}.

First, we identify the issues related to data sharing and trust establishment and maintenance in a virtualized healthcare environment. Second, we define the requirements that the proposed application has to meet. These requirements are mapped to solutions provided with blockchain and smart contracts. Last but not least, we present the proposed architecture, its implementation and evaluation.
To our knowledge, this is the first proof-of-concept designed to enhance trust in a virtualized healthcare environment by utilizing blockchain technology.

We propose three types of evidences for building trust in a virtualized healthcare environment such as evidence of authority, evidence of experience and evidence of competence. Our design uses the public Ethereum blockchain platform, where transactions cost some amount of cryptocurrency. We evaluate the performance of our platform in terms of this cost.

\section{Related work}
Extensive research on the use-cases of blockchain within the health domain has been done in recent years \cite{agbo2019blockchain, HASSELGREN2020104040, mackey2019fit}. The technology is generally proposed as a solution to the data management problems within the health domain, and it is shown as especially well suited for the data sharing problems. These problems relate to challenges with interoperability, security and mobility. The majority of the research on blockchain in healthcare has focused on managing electronic health records \cite{HASSELGREN2020104040}. 

\textbf{MedRec}\cite{7573685} is a proof-of-concept application that relies on the existing data-infrastructure within the healthcare domain. It uses blockchain as a public registry for data sharing and access control of Electronic Medial Records (EMRs). The registry is used to store a simple mapping between a pseudonymous patient identifier, healthcare providers and pointers to EMRs. 
Overall, this architecture allows patients and organizations to locate and access data from a range of providers given a patient's consent. 

\textbf{Ancile}\cite{DAGHER2018283} is a system for controlling access to EMRs, and tries to solve the same problem as MedRec. It improves the previous solutions by including a key-management mechanism for symmetric keys to encrypt the data stored at providers. The system is designed for a permissioned Ethereum-based blockchain, but does not specify the underlying platform further. Permissions for access and participation in the blockchain are governed through a distributed governance mechanism where a pool of voter nodes controls these permissions.

Reference \cite{8210842} defines a set of metrics which can be used for the evaluation of blockchain applications within the health domain. That reference also describes some fundamental principles which should be applied when creating decentralized applications. Although the work is directed towards the American Health Insurance Portability and Accountability Act, the framework can be generalized to a set of specific requirements that can be applied to the European setting. 
References \cite{nichol2016co, funk2018blockchain, capossele2018leveraging} focus on establishing trust in healthcare through blockchain, but they only present conceptual analysis and do not include a practical implementation, design or proof-of-concept. 

\subsection{Using blockchain for trust in healthcare}
To ensure that evidence for a health-workers trust is credible, we can use a blockchain platform for their storage. Blockchains offer data storage, which is immutable and highly distributed, making them easy to access. Blockchain technology has been coined as a key enabling technology for better data-sharing and interoperability within the healthcare industry \cite{gordon2018blockchain}. It can potentially enable patients and healthcare institutions to share, index and control access to data in a fully distributed manner. Blockchain platforms are also, by its nature highly available, and easily accessible by all participants in the blockchain network.

Through the means of smart-contracts, we can create a distributed application running directly on a blockchain platform. These smart-contracts can be used to store the proofs of a health worker's authority, experience and competence while incorporating access control mechanisms that ensure that the published data is credible. The data uploaded via these contracts is immutable, resulting in the proofs of trust that are non-reputable. The non-repudiation property denies any change to the proofs once published, disabling health-workers to alter their proofs fraudulently. Through the use of a public platform, data can be easily available to patients and healthcare institutions, allowing them to validate proofs on-demand.

The Ethereum blockchain is the most popular blockchain platform to incorporate the concept of smart contracts \cite{di20characterizing}. As a consequence, it offers a rich suite of developer tools, enabling rapid prototyping and testing. It, therefore, offers a compelling value-proposition for creating proof-of-concept applications. Although a permissionless blockchain platform in nature, developers stand free to implement their own permission structure within smart contracts to limit how data can be published. Smart contracts on the blockchain can interact with each other, enabling the creation of complex architectures with a rich set of features.

\vspace{-0.3cm}
\subsection{Patient reported outcomes}
One way to measure outcomes from a given treatment and clinical recommendations is through Patient Reported Outcomes (PROs) \cite{weldring2013article}. There are two standardized manners to measure PROs: Patient Reported Outcome Measures (PROMs) and Patient Reported Experience Measures (PREMs) \cite{kingsley2017patient}. They, among other factors, can measure the functional status associated with a treatment or the healthcare which the patients have received. We have chosen the latter to capture the patient experience metrics related to the virtual interaction with the caregiver. These can, for example, be satisfaction rates for patients' experience with their treatment, the health-worker or the virtual setting of the healthcare institution. By creating a link from the healthcare worker to treatment to experience, we can create a model for healthcare worker experience (number of treated cases) and competence (PREMs).  

\section{Used cryptographic components}
VerifyMed uses the Ethereum blockchain \cite{ethyellowpaper} to store data about trust relationships, treatments and evaluations within the health domain. To achieve this, we rely on many cryptographic primitives. Some are used directly, while others are key pieces to understand the underlying workings of the Ethereum blockchain and the tools used to interact with it.
For the overall security of our platform, we followed the design and engineering principles of applied cryptography \cite{schneier2007applied} and we assumed that all security features are inherited from the Ethereum blockchain. While proving the security of some particular and specific relations in our platform is an important issue, in this proof-of-concept stage of the development, it was out of the scope of our work. In that sense, since the purpose of this paper is not to be a tutorial for the cryptographic concepts and primitives that are used in blockchain, we refer an interested reader to some systematization of knowledge publications such as \cite{8865045} and to follow the references there. Yet, we can say that VerifyMed uses the following cryptographic primitives:

\noindent $\bullet$ Cryptographic hash functions (Ethereum uses the NIST approved SHA-3 hash function \cite{dworkin2015sha});

\noindent $\bullet$ Merkle trees (Ethereum uses a generalized form called Modified Merkle Patricia Trees);

\noindent $\bullet$ The ECDSA digital signature scheme \cite{johnson2001elliptic}.

\subsection{Smart contracts in Ethereum}
The main intention of the Ethereum blockchain platform is to enable the creation of general-purpose decentralized applications. The platform can be seen as a state-machine with a set of valid transitions triggered by transactions. Each individual transaction submitted to the blockchain alters the state through the function:
\begin{equation} \label{2:eq:eth-state-transition}
    \sigma_{t+1} = \Upsilon(\sigma_t, T)
\end{equation}
where $\sigma$ (i.e. $\sigma_0, \sigma_1, \ldots$) is the \textit{global state} for the Ethereum blockchain platform, often described as the \textit{world state}. The function $\Upsilon$ is the \textit{Ethereum state transition function}, which produces a new \textit{world state} based on the current \textit{world state} and a \textit{transaction} $T$. To ensure that all nodes participating in the blockchain network can deduce the same \textit{world state} $\sigma$, they must all agree to a fixed ordering of transactions $S = [T_0, T_1, T_2,...]$. Given that such an ordering is shared and agreed upon, all nodes may deduce the same world state by using the transition function over all of these transactions:
\begin{flalign}
    S &= [T_0, T_1, T_2,...] \\
    \sigma_t &= \Upsilon(\Upsilon(\Upsilon(\sigma_0, T_0), T_1), T_2)...
\end{flalign}

The purpose of the blockchain ledger and consensus mechanisms is to allow the nodes in the network to agree to such a transaction order. The ledger follows a general structure where each block contains a set of ordered transactions which are cryptographically bound to the block via a root hash. With the introduction of blocks, we must alter the \textit{world state} update function:
\begin{flalign}
    B_b =& (...,(T_{b1}, T_{b2},...),...) \\
    \sigma_{b} =& \Omega(B_b,\Upsilon(\Upsilon(\Upsilon(\sigma_{b-1}, T_{b0}), T_{b1}), T_{b2})... ) \label{2:eq:block-transition-function}
\end{flalign}
where $\sigma_b$ is the \textit{world state} after block $B_b$ is processed. The block $B_b$ contains the transaction set, along with the remaining data bound to the block. The \textit{Block transition function} $\Omega$ combines the state changes from transactions and the block (e.g. rewards given to the miner of the block) and generates a new \textit{world state} $\sigma_b$.

The main differentiating factor of the Ethereum blockchain platform in comparison to the popular Bitcoin platform is the expressiveness of the \textit{world state} and the ability of users to create smart contracts to utilize this expressiveness. Smart contracts allow users to append their own programs to the blockchain ledger. These programs act like an additional state-machine on top of the existing infrastructure, with their own set of valid transaction types. 

The composition of the Ethereum blockchain platform follows the same fundamental principles of the Bitcoin blockchain platform. However, a fundamental understanding of details related to five different concepts in Ethereum is required and we urge the reader to study reference \cite{ethyellowpaper} for the following Ethereum concepts: 1. Accounts; 2. Smart contracts; 3. Transactions; 4. Costs, and 5. The Ethereum ledger construction.


\vspace{-0.3cm}
\section{Data sharing and trust establishment in virtualized healthcare environment}
A fundamental problem within the health domain is the low capability to share data between healthcare institutions and services. This problem has multiple underlying root-causes, where each can be addressed with a different individual solution. References \cite{HASSELGREN2020104040, MCGHIN201962} have defined four healthcare industry requirements where blockchain can be a significant contributing factor to improvement. Here we explain the same requirements but from a perspective of a healthcare worker.

\begin{enumerate}
    \item \textbf{Interoperability:} Data is not organized in a way that is easily shareable and transferable between institutions. In particular, the data related to the healthcare worker is stored in fragmented data stores where formats and access methods vary from organization to organization. Building applications that can access and integrate all this data together to form an evidence for trust is therefore challenging.
    \item \textbf{Security:} As more digital health data is produced and shared, 
    higher security requirements are imposed. Data providing an evidence for the experience of a healthcare worker is stored in context to patients. Security mechanisms such as access control are therefore patient-centered, making it difficult to access in the context of a health worker without manual intervention.
    \item \textbf{Data sharing:} Due to interoperability and security requirements, it is difficult for patients and healthcare workers to gain access to all their data in a unified view. As healthcare workers change employers, their data documenting their work-history does not follow them. Thus, the evidence of their work-history becomes increasingly fragmented between different organizations over time.
    \item \textbf{Mobility:} 
    A patient traveling between countries, changing services or switching their health domain should be able to transfer his/her data from one health institution to another. Likewise, practitioners should be able to transfer data related to their experience, credentials and practice between institutions. The inability to share the work history of healthcare workers, may limit their ability to move across borders and jurisdictions. Gaining formal certifications and licenses can thus take a long time, reducing the overall efficiency of the healthcare workers and increasing the costs related to recruitment and on-boarding for healthcare institutions.
\end{enumerate}

\subsection{Trust in a virtualized environment}
There is an inherent trust relationship between a patient and healthcare worker in the setting of a physical meeting in a healthcare institution: The patient often trusts that the person in front of him/her in a white coat is an authorised medical professional and the healthcare worker trusts that the patient is whom he/she claims to be, often verified with physical ID \cite{scambler2013system}. The same trust relationship could be extended into a virtualized environment when the patient is talking with a practitioner that the patient already knows from a previous physical setting, and the healthcare worker knows that the patient is who he/she claims to be. Although in a virtualized healthcare environment where the virtual interaction is the first meeting, this same principle cannot be used. Thus, there is a need for establishing such trust relationships in a virtualized healthcare environment. 

To enable trust in a virtualized world, the trustee must be provided with an evidence. This evidence is the ground that justifies a trust relationship between the trustee and the trusted. In the context of the patient-caregiver relationship, we can define the following three major evidences that could enhance trusts:

\begin{enumerate}
    \item \textbf{Evidence of authority:} The healthcare workers must be able to show that they have formal credentials allowing them to practice as healthcare workers. They need a formal license, and their background must be legitimate and approved.
    \item \textbf{Evidence of experience:} The healthcare workers will have the possibility to verify their experience required to deal with the specific health issue of the patient. As specialization increases, this evidence will increasingly be an essential ground for trust. 
    \item \textbf{Evidence of competence:} In addition to being experienced within the medical problem in question, the healthcare workers should be able to show that they have previously delivered positive experiences to other patients. Thus, a metric for patients satisfaction is another crucial evidence.
\end{enumerate}

By making these evidences available to the patients, the grounds for trust between the patient and the healthcare worker can be established. However, designing such a solution is not trivial due to the major requirements defined within the health domain: interoperability, security, data sharing and mobility. These requirements make it challenging to create an application that works on top of the existing organizational structure where data is fragmented over different organizations with a diverse set of formats. 


Making data about healthcare workers' authority, experience and competence transparent, available and immutable can be perceived as a privacy issue for healthcare workers. However, this structure also has some significant advantages for the healthcare worker. Due to the availability of data, turnover, on-boarding and mobility processes can be simplified due to the ability of employers to perform efficient background checks related to their profession. They can also have better visibility, providing a major incentive to provide better care. 

\section{Description of the architecture}
\begin{figure}
    \centering
    \hspace{-0.8cm}\includegraphics[width=0.8\textwidth]{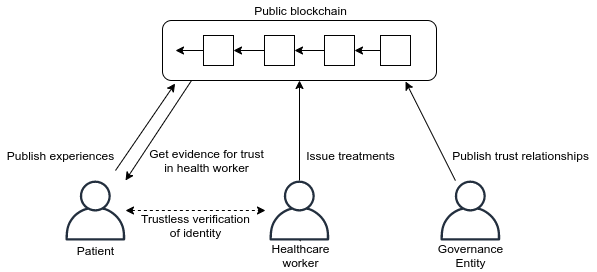}
    \caption{Interacting with the blockchain to gain trust in a health worker}
    \label{4:fig:trust-architecture}
    \vspace{-0.25cm}
\end{figure}

Our proposed system architecture is designed to store Evidences of Authority, Evidences of experience and Evidences of competence on the public Ethereum blockchain platform. To enable these evidences to hold any legitimacy, the application incorporates a concept of governance, where a set of stakeholders cooperate to create a trusted environment on the blockchain via smart contracts. Patients use this trusted environment to gain evidence for trust in a health worker, and publish their own experiences once the patient and health worker interaction is completed. While the high-level view, as shown in Figure \ref{4:fig:trust-architecture}, is simple, the underlying system design is of high complexity. 
The first contributor to increased complexity is the real-world trust relationships within the healthcare system. While our top level model depicts a single \textit{governance entity}, no such entity exists in the real world. The trust relationships within the healthcare industry include a broad set of different organizational entities. These entities hold specific responsibilities, and they can only together create overall trust in the healthcare system. Our system architecture includes multiple organizational entities, and we capture the trust relationships between them on the blockchain. 

Other requirements relevant to our platform include patient privacy, prevention of fraudulent patient evaluations and scalability considerations. These quality attributes can only be addressed through architectural choices, furthermore contributing to complexity. After describing our overall system architecture and our choices, we will break this down into procedures and subsections to show how the architecture addresses these requirements.

\subsection{Modeling evidence for trust}
\subsubsection{Evidence of authority} \ \ 

The first evidence for trust in a healthcare worker is the evidence of authority. This evidence consists of the formal credentials which allow the healthcare professional to practice. By providing this evidence on a blockchain, patients or any other interested stakeholder can access it freely. If the patient can confirm the link between a healthcare worker and the evidence of authority on the blockchain, he/she should be able to trust that the healthcare worker has formal authorization. In practice, we choose to model the evidence of authority as two different statements which the healthcare worker wants to prove:
\begin{enumerate}
    \item The healthcare worker is currently in possession of a valid License for Health Personnel in the area he or she operated and is thus formally qualified to practice.
    \item The healthcare worker is formally associated with an authorized healthcare facility.
\end{enumerate}
Both of these statements cannot be fulfilled by the healthcare worker alone. They are instead statements of trust from other organizational entities that are deemed trusted themselves. This structure of entities and their trust relationships quickly serves as the foundation of trust in the system. We  define the following stakeholders that create one hierarchy of trust:
\begin{itemize}
    \item \textbf{Authorities} are top-level healthcare authorities responsible for the formal authorization of healthcare institutions, educational facilities and other organizations who provide healthcare related services. Organizations with such authorities are usually the national health directorates. These organizations organize themselves via a distributed governance protocol.
    \item \textbf{License Issuers} are organizations that are responsible for the formal authorization of healthcare workers. They are responsible for background checking of the applicant and use their documented experience and performance to decide if the healthcare worker is fit to hold a License. If that is the case, they choose to issue such a license and thus establish a trust relationship with the healthcare worker. Such organizations are often units within a national health directorate.
    \item \textbf{License Providers} are authorized healthcare facilities responsible for the practice of the healthcare worker on a day-to-day basis. These facilities are under continuous evaluation by the authorities and have to ensure the competence of their associated healthcare workers. Such organizations can include hospitals or clinics.
    \item \textbf{Treatment Providers} are health service providers who are responsible for facilitating the interactions between patients and healthcare workers. They hold the main responsibility for authenticating patients and for storing data related to the interaction. Such stakeholders may be similar to license providers, like hospitals or clinics. It can also include services such as e-health platforms and secondary consultation services.
    \item \textbf{Licenses} are the components that represent the healthcare workers within our trust model. A license can only be created by a license issuer, and it is tied to credentials (keys) in possession of the health worker. Once issued, it may be transferred between License Providers, License Issuers and associated with additional Treatment Providers if these stakeholders agree to these movements.
\end{itemize}

\shorten{
\begin{figure}
    \centering
    \hspace{-0.5cm}\includegraphics[width=0.5\textwidth]{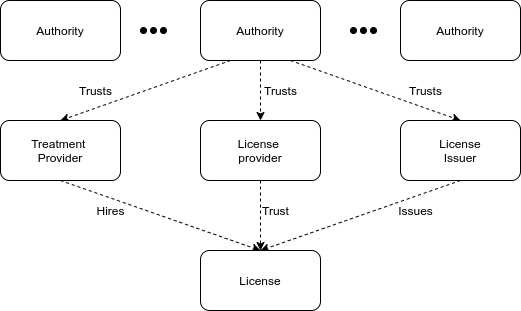}
    \caption{A trust model for stakeholders in the healthcare industry}
    \label{4:fig:trust-relations}
\end{figure}
} 

Together, these stakeholders interact and build a complete trust hierarchy. 
This model is captured via smart contracts deployed to the blockchain ledger, storing data about stakeholders and their trust relationships.

\shorten{
The top level is our authorities, where the smart contract creator start as an authority as a bootstrap. These authorities organize themselves via a distributed governance protocol. We propose to use simple majority voting for this purpose, but implementations stand free to choose any protocol of their liking.

Authorities may in turn place their trust in License Issuers, License Providers and Treatment Providers. These trust relationships should be justified in real-world decisions and processes to deem them eligible to hold such a responsibility. If an authority deems their trusted stakeholders to be unfit of their trust, they can choose to revoke this trust relation.

Validating the two statements for \textcolor{red}{evidence} of authority is done by asserting the current state in the trust hierarchy as represented on the blockchain. Checking if a healthcare worker has a valid License for Health Personnel is done by inspecting their license representation and validate that a trust relationship exist from a License Issuer. The license issuer must in addition be trusted by an authority. A similar process is performed to validate the association with a trusted License provider.
} 

\subsubsection{Evidence of experience} \ \ 

The second evidence for trust in healthcare workers is their experience. Depending on the context in which the patient meets a healthcare worker, experience within a relevant field may be of high importance to ensure that the healthcare worker can deliver the care required. The metrics for experience come in either qualitative or quantitative forms. The qualitative evidence can be conveyed through certifications, while the quantitative evidence can be deduced from metrics such as the number of a specific treatment performed by the healthcare worker or the number of specific problems addressed. To model the evidence of experience, we choose to focus on quantitative metrics.

\begin{figure}
    \centering
    \includegraphics[width=0.8\textwidth]{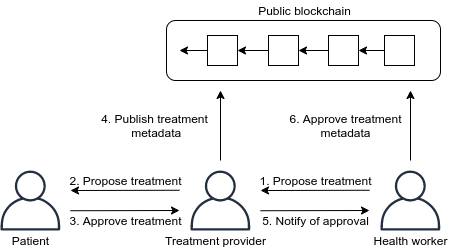}
    \caption{A model for generating evidences for the experience of health workers}
    \label{4:fig:proof-of-experience-model}
\end{figure}

The goal of our model is to expose the number of treatments performed by the healthcare worker to the patient. Evidence of authority is created by a formal model for creating an evidence. In contrast, evidence of experience is generated through patient and healthcare worker interactions. Each new interaction resulting in a treatment thus forms evidence for future patients who want to interact with the healthcare worker. Figure \ref{4:fig:proof-of-experience-model} shows our model for publishing treatment information on the blockchain. During the patient and healthcare worker interaction, the treatment provider is responsible for conveying information about treatments recommended by a healthcare worker. Once approved by a patient the full content of the treatment is stored at the treatment provider. Metadata about the treatment is published to the blockchain, which is in turn approved publicly by the healthcare worker, thus forming a public link from the healthcare worker to the treatment. Over time, this process will generate a public log capturing metadata about treatments performed by a health worker, which serves as the evidence of experience.

\subsubsection{Evidence of competence} \ \ 

\begin{figure}
    \centering
    \includegraphics[width=0.8\textwidth]{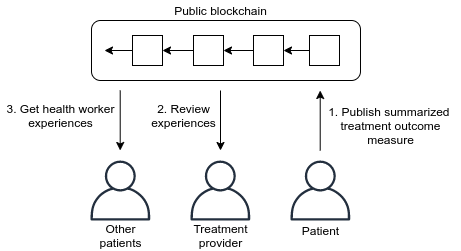}
    \caption{A model for generating evidences for the competence of health workers}
    \label{4:fig:proof-of-comperance-model}
\end{figure}

While a quantitative metric like a number of treatments can be evidence for experience, it does not represent the quality of these treatments. Patient Reported Experience Measures (PREMs) is a standardized way to measure the outcome of an encounter. By summarizing these outcomes into qualitative metrics which is published on the blockchain, we can measure the quality of a treatment. This general process is shown in Figure \ref{4:fig:proof-of-comperance-model}, where the patient interacts directly with the blockchain to publish a summarized outcome measure related to a treatment they have gone through. Since these treatments are linked to a healthcare worker, we can use them as a proxy for evaluating their competence. As a log of treatment metadata with corresponding outcome measures is built on the blockchain, it serves as evidence of competence.


\subsubsection{Access control in smart contracts} \ \ 

\shorten{
Our architecture is designed to use a public and permissionless blockchain platform. Such platforms allow for any user to submit a transaction targeted for our smart contracts. To preserve the legitimacy of trust relationships, treatments and evaluations published via our blockchain applications, such transactions must be handled by access control schemes. These schemes must only allow authorized senders to change the state within our Blockchain application. This behaviour can be achieved by implementing access control logic within the smart contracts themselves, which is our practiced method in the architecture.
}

The implemented access control scheme can be described as a Role Based Access Control (RBAC) scheme where accounts interacting with the blockchain must hold a certain role within our distributed application to perform such an action. Examples of access control policies in the blockchain component of our architecture include: \textbf{1.} Only existing authorities may interact with the distributed governance protocol; \textbf{2.} Only existing authorities may add trust in a treatment provider, license provider or license issuer; \textbf{3.}  Only treatment providers trusted by an authority may add treatments; and \textbf{4.} Evaluations can only be created by the patient who is the subject in a treatment.

\shorten{
These policies are part of the internal contract logic and enforced though interaction between the contracts. An example of this process is shown in Figure \ref{4:fig:component:create-treatment-dDapp}, where the treatment contract checks if the sender of the transaction is indeed a trusted treatment provider. We notice how this check is propagated all the way to the authority contract. Thus, if a authority is removed, any trusted descendants will loose their authority to create treatments immediately. 

\begin{figure*}
    \centering
    \includegraphics[width=\textwidth]{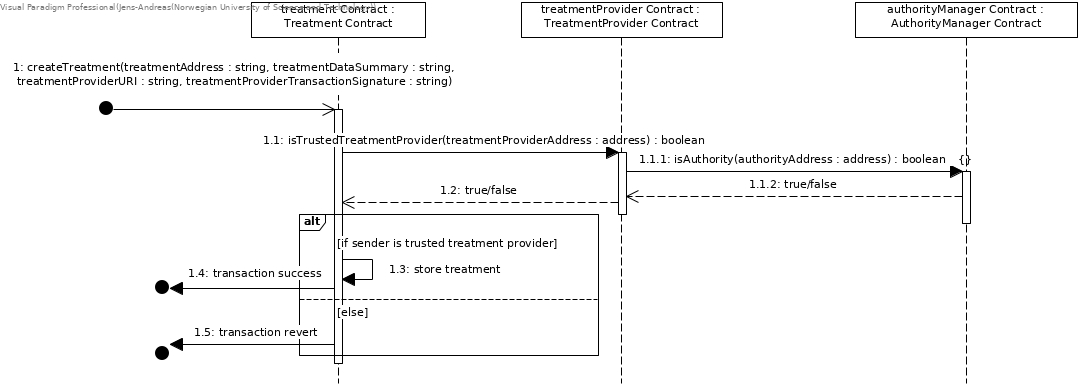}
    \caption{A sequence diagram showing authentication flow when submitting a create treatment transaction.}
    \label{4:fig:component:create-treatment-dDapp}
\end{figure*}

} 

\section{Implementation details}
A fully working proof-of-concept application was developed for assembling metrics, finding faults with the architecture, and for testing stakeholder workflows. During the application development process, we tried to keep usability for administrators in mind, where we tried to make the process of administering as easy to run as possible. This will allow further development of the application to be easy, and allows third parties to easily test and set up the application. All the code for the software application is in a github repository\footnote{\url{https://github.com/jarensaa/transparent-healthcare}}. 

\begin{figure}
    \centering
    \includegraphics[width=0.8\textwidth]{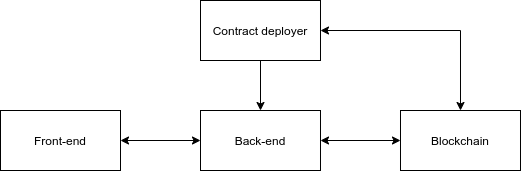}
    \caption{An overview of the run-time presence for our implemented services}
    \label{c:fig:runtime-presence}
\end{figure}

The full application is created as four independent services. These services together simulate the architectures shown in Figure \ref{4:fig:trust-architecture}, \ref{4:fig:proof-of-experience-model} and \ref{4:fig:proof-of-comperance-model}. Figure \ref{c:fig:runtime-presence} shows how these services interact during runtime. The contract-deployer service is a short-lived service responsible for deploying the smart contracts to the blockchain and export the keys used for their deployment.


\subsection{Setup with Docker}
A \texttt{docker-compose.yml} file is provided in the root of the project. This file is designed to automatically build and start all services in the correct order. Running \texttt{docker-compose up} in the project root should be sufficient.

\shorten{
\subsection{Starting services manually}
When developing on the application, you probably want to run the services manually. To do this, follow this procedure: 
\begin{enumerate}
    \item Open a terminal window and navigate to the \texttt{ganache} folder.
    \item Run \texttt{yarn start}
    \item Open another terminal window and navigate to the \texttt{contracts} folder.
    \item Run \texttt{yarn start}
    \item Open another terminal window and navigate to the \texttt{providerService} folder.
    \item Run \texttt{./gradlew bootRun}
    \item Open another terminal window and navigate to the \texttt{webapp} folder.
    \item Run \texttt{yarn start}
\end{enumerate}
} 

\begin{figure*}
    \centering
    \includegraphics[width=0.8\textwidth]{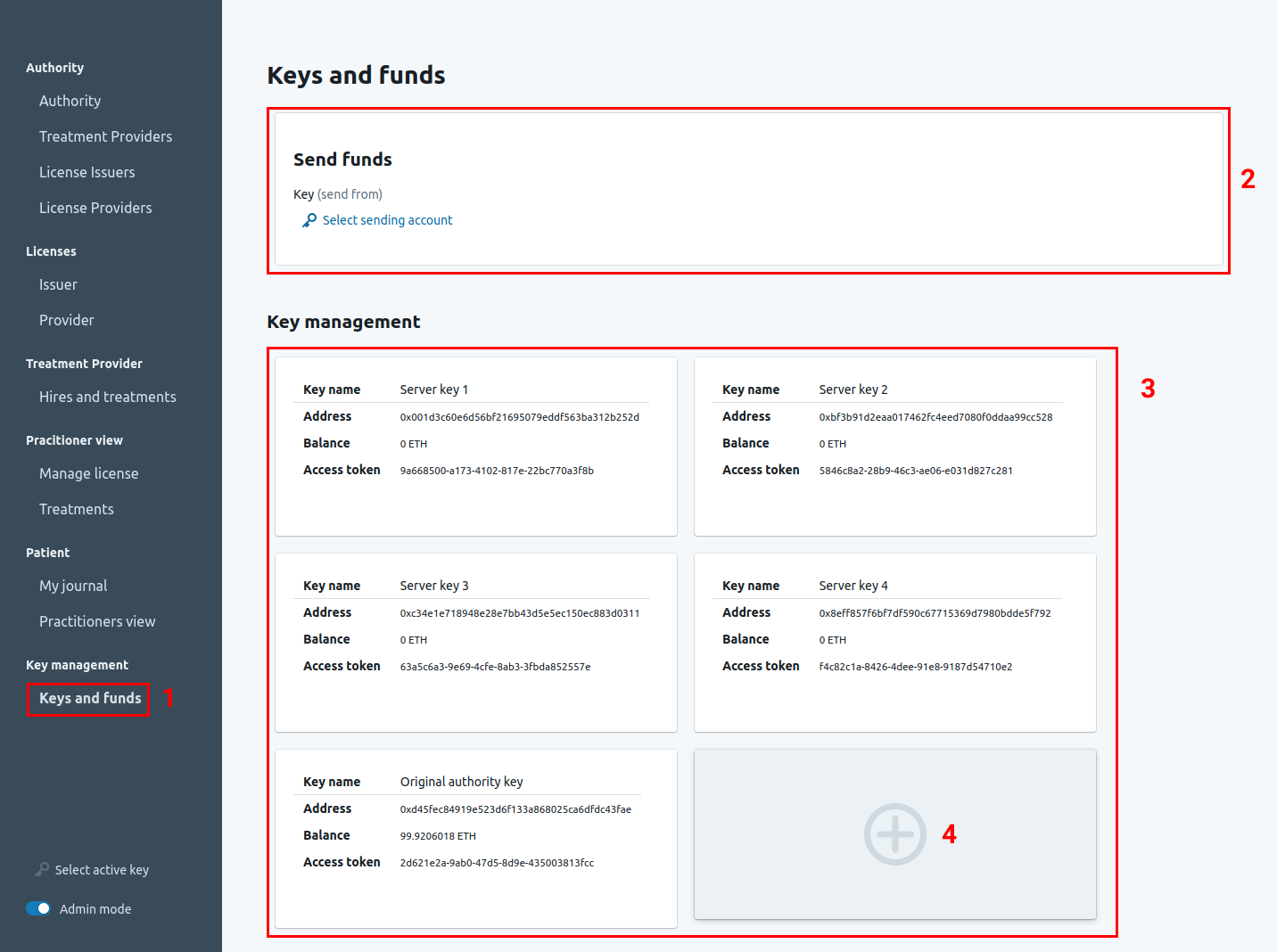}
    \caption{Overview of the key management panel}
    \label{b:fig:key-panel}
\end{figure*}

\subsection{User interface}
Once the platform is up and running it offers a user interface with the following sections:
\begin{enumerate}
    \item Section for navigating to the panels relevant to the authority stakeholder;
    \item Section for navigating to the panels relevant to the License Provider and license issuer stakeholder;
    \item Section for navigating to the panels relevant to the Treatment Provider stakeholder;
    \item Section for navigating to the panels relevant to the healthcare worker stakeholder;
    \item Section for navigating to the panels relevant to the patient stakeholder;
    \item Section for navigating to the panels relevant to key management, relevant to all stakeholders;
    \item Selection button for selecting the current active Ethereum keypair (ECDSA keypair) to be used for actions in the UI, relevant for all stakeholders;
    \item Toggle for \textit{admin mode}: This gives access to an account which is the first default authority, thus, giving a baseline allowing the user to expand the hierarchy from there. This account has a initial balance of 100ETH, which can be sent to other accounts so they are able to create transactions.
\end{enumerate}

\subsection{Key management}
The key management is performed via a panel shown in Figure \ref{b:fig:key-panel}. This panel allows users to create and view keypairs. These keys are either stored on the server or locally, it depends on the intent. One can also use the panel to send Ether from one account to another. The panel has the following sections:
\begin{enumerate}
    \item The button to click to access the key management panel.
    \item The panel to send Ether from one account to another.
    \item The section to view current keys of all formats. This shows fields such as address and balance. If a local key is shown, the private key will be used. If present, an access token to use the key on the backend is shown.
    \item A card which can be clicked to create new keys with a selection of types.
\end{enumerate}

\shorten{
\subsection{Sending Ether}
Section 2 in Figure \ref{b:fig:key-panel} allows all stakeholders to send funds between the keys they have access to. Figure \ref{b:fig:key-send} shows how the interaction with the panel, where the following steps are taken to send funds:

\begin{enumerate}
    \item Select the key to send funds from
    \item Select the amount of Ether to send
    \item Select the key to send funds to
    \item Click to send the send funds transaction to the blockchain.
\end{enumerate}

This will create a new send transaction on the blockchain platform. Once the transaction is added to the ledger, the user may refresh the page to see the new balances. 

\begin{figure*}
    \centering
    \includegraphics[width=0.7\textwidth]{key-send.png}
    \caption{The send funds panel in use}
    \label{b:fig:key-send}
\end{figure*}
} 

\subsection{Initial results of simulated use of the platform}
We have performed numerous simulated runs for different workflows of the platform. The simulations were performed with a single-node Ethereum network configured to simulate the real-world behaviour of the public Ethereum network. Configuration parameters for the blockchain were: Block gas limit: 9.991.391; Block generation time: 20s.

The simulations showed the following costs (in gas denominations) for different smart contract invocations: Add another address as a authority - 179013; Remove an authority - 149040; Vote on a proposal to add or remove a authority - 73686; Enact a proposal to add or remove a authority - 64297; Trying to enact a proposal without a majority vote in place - 28045; Enact proposal to remove an authority - 45332; Add trust in a registered treatment provider - 93707; Remove trust in a registered treatment provider - 22909; Add trust in a registered license issuer - 48863; Remove trust in a registered license issuer - 18829; Add trust in a registered license provider - 48906; Remove trust in a registered license provider - 15965; Register address as a treatment provider - 85959; Create a new treatment - 200118; Register as license issuer - 71059; Issue a new license to address - 88538; Approve movement of license to a new license issuer - 23040; Register as license provider - 86036; Approve movement of license to a new license provider - 38019; Propose movement of license to a new license provider - 46059; Propose license provider movement - 46092; Approve published treatment for a given patient - 102721; Submitting an evaluation - 143669.

\begin{figure}
    \centering
    \includegraphics[width=0.65\textwidth]{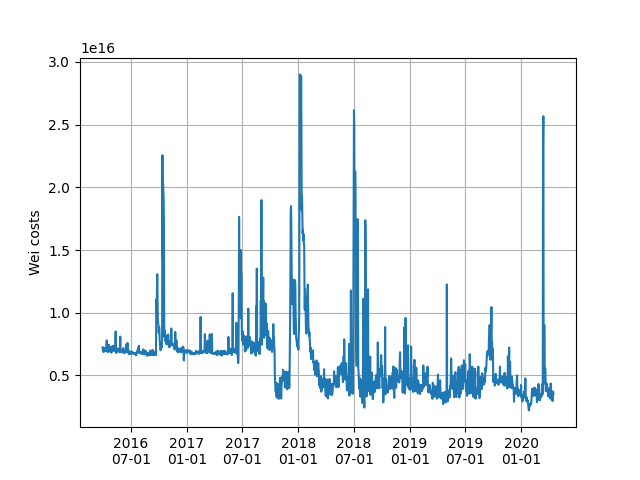}
    \caption{Cost in wei for creating a treatment and getting it approved by a health worker.}
    \label{6:plt:wei:treatment}
\end{figure}

\begin{figure}
    \centering
    \includegraphics[width=0.65\textwidth]{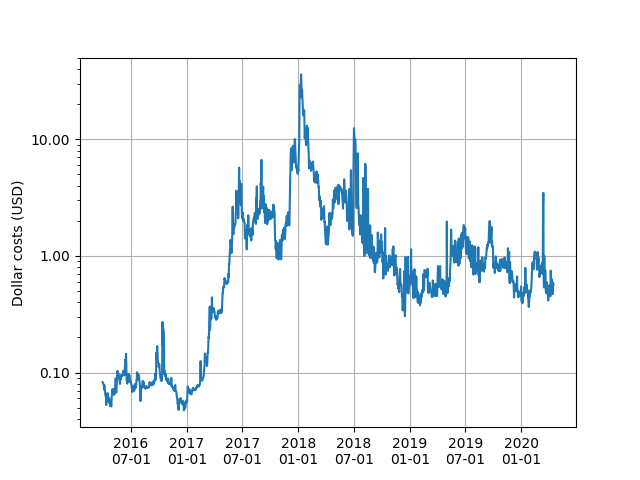}
    \caption{Cost in USD for creating a treatment and getting it approved by a health worker.}
    \label{6:plt:usd:treatment}
\end{figure}

\begin{figure}
    \centering
    \includegraphics[width=0.65\textwidth]{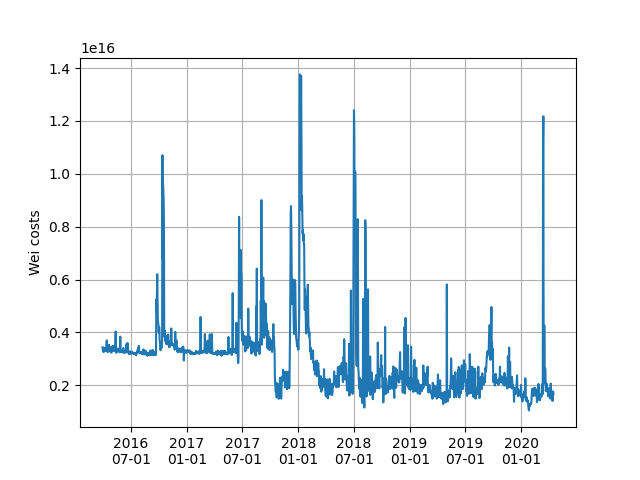}
    \caption{Cost in wei for evaluating a treatment}
    \label{6:plt:wei:evaluate}
\end{figure}

\begin{figure}
    \centering
    \includegraphics[width=0.65\textwidth]{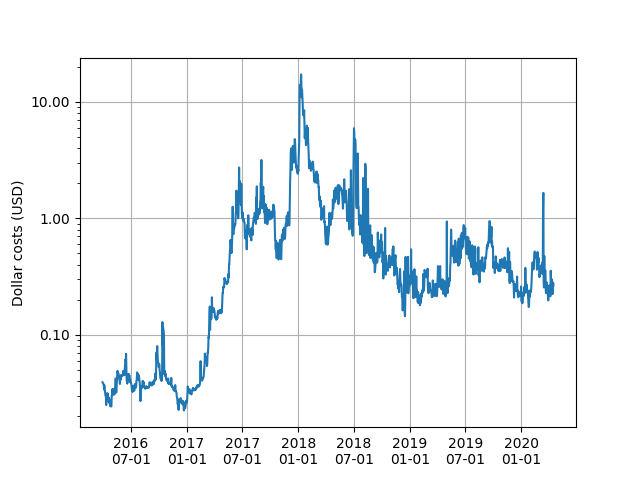}
    \caption{Cost in USD for evaluating a treatment}
    \label{6:plt:usd:evaluate}
\end{figure}

Combined with historical data of the price of Ether vs. USD, in Figures \ref{6:plt:wei:treatment}, \ref{6:plt:usd:treatment}, \ref{6:plt:wei:evaluate} and \ref{6:plt:usd:evaluate}, we show the simulated costs for several smart contracts such as creating a treatment or evaluation of a treatment. In Figure \ref{6:plt:usd:treatment}, we show that the current cost for creating a treatment and getting it approved is around 1 USD, and the cost for evaluating a treatment, as shown in Figure \ref{6:plt:usd:evaluate}, would be around 0.5 USD. However, these prices may increase dramatically if network congestion reaches similar levels as in January 2018, when a dramatic cost increase was observed.

\shorten{
\begin{figure}
    \centering
    \includegraphics[width=0.65\textwidth]{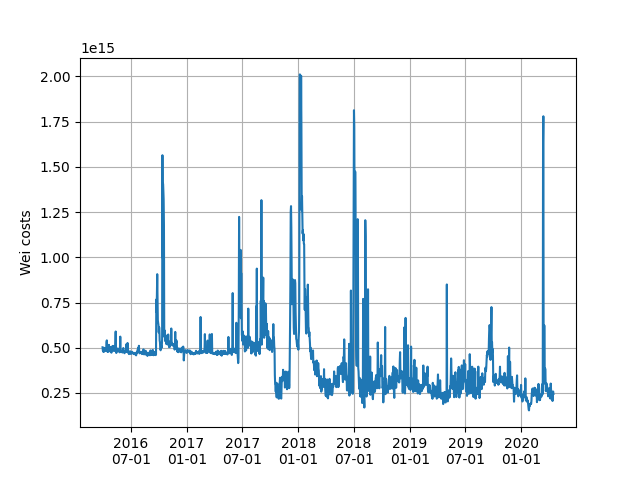}
    \caption{Cost in wei for sending Ether}
    \label{6:plt:wei:send}
\end{figure}

\begin{figure}
    \centering
    \includegraphics[width=0.45\textwidth]{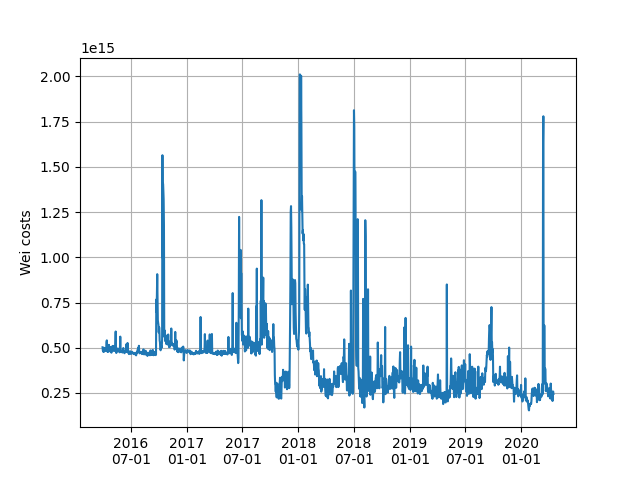}
    \caption{Cost in USD for sending Ether}
    \label{6:plt:usd:send}
\end{figure}
} 

\section{Conclusions and future fork}
We presented the design rationale, modelling and implementation of VerifyMed - a robust blockchain platform for transparent trust in a healthcare domain. To our knowledge, this is one of the first blockchain solutions addressing this specific problem. It is based on the Ethereum blockchain. Our platform is released as an open source code in github. The open source includes also user guides for setting up and running the platform. 

We envision three user entities for this trust enhancing platform: governance entities, healthcare workers and patients. For each of these entities, the platform offers easy and intuitive user interfaces.

We have performed numerous simulated use case scenarios with the platform and showed the modest cost of the platform services for an extended simulated period of four years.

\textit{Future work:} Our platform in the future updates will enrich the current trust model by including more trust requirements such as 1. The caregiver must trust that the patient exists; 2. The caregiver must trust the authenticity of the data that the patient is willing to share and 3. A third party (e.g. a insurance company) must be able to trust the patients claim that care provision has taken place.

\bibliographystyle{acm}
\bibliography{reference.bib}

\begin{thebibliography}{10}

\bibitem{agbo2019blockchain}
{\sc Agbo, C.~C., Mahmoud, Q.~H., and Eklund, J.~M.}
\newblock Blockchain technology in healthcare: a systematic review.
\newblock In {\em Healthcare\/} (2019), vol.~7, Multidisciplinary Digital
  Publishing Institute, p.~56.

\bibitem{7573685}
{\sc {Azaria}, A., {Ekblaw}, A., {Vieira}, T., and {Lippman}, A.}
\newblock Medrec: Using blockchain for medical data access and permission
  management.
\newblock In {\em 2016 2nd International Conference on Open and Big Data
  (OBD)\/} (Aug 2016), pp.~25--30.

\bibitem{capossele2018leveraging}
{\sc Capossele, A., Gaglione, A., Nati, M., Conti, M., Lazzeretti, R., and
  Missier, P.}
\newblock Leveraging blockchain to enable smart-health applications.
\newblock In {\em 2018 IEEE 4th International Forum on Research and Technology
  for Society and Industry (RTSI)\/} (2018), IEEE, pp.~1--6.

\bibitem{DAGHER2018283}
{\sc Dagher, G.~G., Mohler, J., Milojkovic, M., and Marella, P.~B.}
\newblock Ancile: Privacy-preserving framework for access control and
  interoperability of electronic health records using blockchain technology.
\newblock {\em Sustainable Cities and Society 39\/} (2018), 283 -- 297.

\bibitem{di20characterizing}
{\sc Di~Angelo, M., and Salzer, G.}
\newblock {Characterizing Types of Smart Contracts in the Ethereum Landscape}.
\newblock In {\em Proc. 4th Workshop on Trusted Smart Contracts, Financial
  Cryptography 2020\/} (2020), Springer.

\bibitem{dworkin2015sha}
{\sc Dworkin, M.~J.}
\newblock {SHA-3 standard: Permutation-based hash and extendable-output
  functions}.
\newblock Tech. rep., {Nationa Institute of Science and technology (NIST)},
  2015.

\bibitem{funk2018blockchain}
{\sc Funk, E., Riddell, J., Ankel, F., and Cabrera, D.}
\newblock Blockchain technology: A data framework to improve validity, trust,
  and accountability of information exchange in health professions education.
\newblock {\em Academic Medicine 93}, 12 (2018), 1791--1794.

\bibitem{gordon2018blockchain}
{\sc Gordon, W.~J., and Catalini, C.}
\newblock Blockchain technology for healthcare: facilitating the transition to
  patient-driven interoperability.
\newblock {\em Computational and structural biotechnology journal 16\/} (2018),
  224--230.

\bibitem{HASSELGREN2020104040}
{\sc Hasselgren, A., Kralevska, K., Gligoroski, D., Pedersen, S.~A., and
  Faxvaag, A.}
\newblock Blockchain in healthcare and health sciences—a scoping review.
\newblock {\em International Journal of Medical Informatics 134\/} (2020),
  104040.

\bibitem{johnson2001elliptic}
{\sc Johnson, D., Menezes, A., and Vanstone, S.}
\newblock The elliptic curve digital signature algorithm (ecdsa).
\newblock {\em International journal of information security 1}, 1 (2001),
  36--63.

\bibitem{kingsley2017patient}
{\sc Kingsley, C., and Patel, S.}
\newblock Patient-reported outcome measures and patient-reported experience
  measures.
\newblock {\em Bja Education 17}, 4 (2017), 137--144.

\bibitem{mackey2019fit}
{\sc Mackey, T.~K., Kuo, T.-T., Gummadi, B., Clauson, K.~A., Church, G.,
  Grishin, D., Obbad, K., Barkovich, R., and Palombini, M.}
\newblock ‘fit-for-purpose?’--challenges and opportunities for applications
  of blockchain technology in the future of healthcare.
\newblock {\em BMC medicine 17}, 1 (2019), 68.

\bibitem{MCGHIN201962}
{\sc McGhin, T., Choo, K.-K.~R., Liu, C.~Z., and He, D.}
\newblock Blockchain in healthcare applications: Research challenges and
  opportunities.
\newblock {\em Journal of Network and Computer Applications 135\/} (2019), 62
  -- 75.

\bibitem{nichol2016co}
{\sc Nichol, P.~B., and Brandt, J.}
\newblock Co-creation of trust for healthcare: The cryptocitizen framework for
  interoperability with blockchain.
\newblock {\em Research Proposal. ResearchGate\/} (2016).

\bibitem{8865045}
{\sc {Raikwar}, M., {Gligoroski}, D., and {Kralevska}, K.}
\newblock Sok of used cryptography in blockchain.
\newblock {\em IEEE Access 7\/} (2019), 148550--148575.

\bibitem{raikwar2020trends}
{\sc Raikwar, M., Gligoroski, D., and Velinov, G.}
\newblock {Trends in Development of Databases and Blockchain}.
\newblock arXiv:2003.05687, 2020.
\newblock \url{https://arxiv.org/pdf/2003.05687.pdf}.

\bibitem{scambler2013system}
{\sc Scambler, G., and Britten, N.}
\newblock System, lifeworld and doctor--patient interaction: Issues of trust in
  a changing world.
\newblock In {\em Habermas, critical theory and health}. Routledge, 2013,
  pp.~53--75.

\bibitem{schneier2007applied}
{\sc Schneier, B.}
\newblock {\em Applied cryptography: protocols, algorithms, and source code in
  C}.
\newblock John Wiley \& sons, 2007.

\bibitem{Thuemmler2017}
{\sc Thuemmler, C., and Bai, C.}
\newblock {\em Health 4.0: Application of Industry 4.0 Design Principles in
  Future Asthma Management}.
\newblock Springer International Publishing, Cham, 2017, pp.~23--37.

\bibitem{weldring2013article}
{\sc Weldring, T., and Smith, S.~M.}
\newblock Article commentary: Patient-reported outcomes (pros) and
  patient-reported outcome measures (proms).
\newblock {\em Health services insights 6\/} (2013), HSI--S11093.

\bibitem{ethyellowpaper}
{\sc Wood, G.}
\newblock Ethereum yellow paper.
\newblock {\em Internet: https://github. com/ethereum/yellowpaper, [version
  7e819ec - 2019-10-20]\/} (2014).

\bibitem{8210842}
{\sc {Zhang}, P., {Walker}, M.~A., {White}, J., {Schmidt}, D.~C., and {Lenz},
  G.}
\newblock Metrics for assessing blockchain-based healthcare decentralized apps.
\newblock In {\em 2017 IEEE 19th International Conference on e-Health
  Networking, Applications and Services (Healthcom)\/} (Oct 2017), pp.~1--4.

\end{thebibliography}

\end{document}